\documentclass[a4paper,openright,10pt]{article}
\usepackage{fancyhdr}
\usepackage{graphicx} 
\usepackage{amsfonts}
\usepackage{amsmath,amsthm,amssymb}
\usepackage{mathptmx}

\usepackage[utf8]{inputenc}
\usepackage{dsfont}
\usepackage[english]{babel}
\usepackage{cite}
\usepackage{subfigure} 
\usepackage[hyperindex=true,breaklinks=true,colorlinks=true,linkcolor=blue,citecolor=blue]{hyperref}

\numberwithin{equation}{section}

\title{\bf Polar coherent states in bilayer graphene under a constant uniform magnetic field}

\author{
D.I. Mart\'inez-Moreno$^1$\footnote{dmartinez@fis.cinvestav.mx, ORCID: \href{http://orcid.org/0000-0001-9583-4873}{0000-0001-9583-4873}}, \ 
J. Negro$^2$\footnote{jnegro@fta.uva.es, ORCID: \href{http://orcid.org/0000-0002-0847-6420}{0000-0002-0847-6420}}, \ and 
L.M. Nieto$^2$\footnote{luismiguel.nieto.calzada@uva.es, ORCID: \href{http://orcid.org/0000-0002-2849-2647}{0000-0002-2849-2647}}
\medskip
\\ 
\small
\noindent
$^1$\,Departamento de F\'{\i}sica, CINVESTAV, AP 14-740, 07000 Ciudad de M\'exico, M\'exico
\\
\small
\noindent
$^2$\,Departamento de F\'{\i}sica Te\'orica, At\'omica y
\'Optica, Universidad de Valladolid,  47011 Valladolid, Spain
}

\begin{document}

\maketitle

\begin{abstract}
Symmetries associated with the Hamiltonian describing bilayer graphene  subjected to a constant magnetic field perpendicular to the plane of the bilayer are calculated using polar coordinates. These symmetries are then applied to explain some fundamental properties, such as the spectrum and the integer pseudo-spin character of the eigenfunctions.
The probability and current densities of the bilayer Hamiltonian have  also been calculated in polar coordinates and shown to be gauge invariant and scalar under generalized rotations.
We also define  appropriate coherent states of this system as eigenfunctions,  with complex eigenvalues, of a suitable chose annihilation operator. 
In this framework,  symmetries are also useful to show the meaning of the complex eigenvalue in terms of expected values. 
The local  current density of these coherent states is shown to exhibit a kind of  radial component interference effect, something that has gone unnoticed until now.  
Some of these results that have just been exposed are graphically illustrated throughout the manuscript.
\end{abstract}

\section{Introduction}

Since the experimental discovery of graphene \cite{NovoselovScience} many theoretical and experimental studies have been carried out on this material as well as on other models of two-dimensional physical systems \cite{K2012,castro}. 
The present work will focus on a particular type of such material, the so-called bilayer graphene, which consists of two layers of coupled carbon atoms, each with a hexagonal crystalline structure.
Depending on how these layers are arranged, bilayer graphene can have different properties, for example, if the layers are rotated through a certain angle known as {\it magic angle}, bilayer graphene displays unconventional superconductivity and Mott-like insulating states \cite{{CFF2018,C2018}}. However, in the present work we will consider the most common arrangement, known as Bernal stacking \cite{MF2006}. 

It is known that the electronic band structure of bilayer graphene can be analyzed using the {\it tight-binding model} \cite{MCMK2013,MAF2007} through which it is possible to predict the existence of four energy bands, two conduction bands and two valence bands. 
Near   certain points of symmetry, one of these conduction bands touches a valence band at zero energy and this implies that at low energies, the electrons in the bilayer graphene behave as chiral quasiparticles of effective mass $m^{*}$ with a parabolic dispersion relation $E=\pm p^{2}/2m^{*}$, unlike the lineal relationship that occurs in the graphene monolayer.
In a recent work, the interaction of these bilayer graphene chiral electrons with different \cite{FGO20} magnetic field profiles has been studied.
In particular, for the simplest case in which the bilayer graphene is placed in a constant magnetic field perpendicular to its surface, coherent states have been constructed to describes this quantum system through a semiclassical approach \cite{FM20}.

The main objective of this work is to complete these studies by analyzing the bilayer graphene under a constant magnetic field  using a symmetric gauge and polar coordinates,   instead of the Landau gauge that has been used in previous studies. 
In particular, the current densities of bilayer graphene will be derived in detail and calculated for both the eigenstates of the system and the radially symmetric coherent states, which are also constructed in the current work. 

The organization of this work is as follows. In Section~\ref{modelobilayer} the mathematical model is presented and the matrix Hamiltonian is decoupled into two fourth-order scalar Hamiltonians in the derivatives.
In Section~\ref{simetrias} we study the symmetries of these two scalar Hamiltonians and use them to find their spectra.
In Section~\ref{spectrumbilayer} we found the spectrum and study the properties of the bilayer Hamiltonian using the results of the scalar Hamiltonians. 
Furthermore, in this section the correct expressions of the current densities for bilayer graphene  when subjected to a constant magnetic field are derived. 
Then, the radially symmetric coherent states of bilayer graphene are built as eigenstates of an appropriate annihilation operator with a complex eigenvalue $\alpha^{2}$ and some properties of these states, including their current densities, are studied in Section~\ref{coherentbilayer}. Finally, the paper ends by outlining the main conclusions in Section~\ref{conclusion}.

\section{The bilayer Hamiltonian}\label{modelobilayer}

As is well known in the literature \cite{K2012,MCMK2013}, the Hamiltonian that describes the electronic behavior in a graphene bilayer structure located in the $x{-}y$ plane is given by
\begin{equation}
{\bf H} = \frac{1}{2m^*}
\left(\begin{array}{cc}
0 & (p_{x} - ip_{y})^2 \\ 
(p_{x} + ip_{y})^2 & 0
\end{array}
\right), 
\label{eq.bilayer-hamiltonian}
\end{equation}
where $\vec{p}=-i\hbar\nabla$ and $m^*$ is the effective mass of the chiral electrons, such that $m^*\approx$ 0.054$m_{e}$, with $m_{e}$ the  mass of the electron. 

As we have already indicated, in order to describe the radial and rotational symmetries of the interaction of bilayer graphene electrons, with charge $q=-e$ and effective mass $m^*$, with a constant homogeneous magnetic field perpendicular to the bilayer surface ($\vec{B}= B_{0}\vec{k}$), the symmetric gauge will be used  for the vector potential, which is given by
\begin{equation}
\vec{A}=\left( -\dfrac{B_{0}}{2}y, \dfrac{B_{0}}{2}x, 0 \right), 
\label{eq.vector-potential} 
\end{equation} 
such that $\vec{B}= \nabla \times \vec{A}= (0, 0, B_{0})$.
Using the minimal coupling recipe ($\vec{p}\rightarrow\vec{p}-q\vec{A}/c$) to insert this vector potential  into the Hamiltonian \eqref{eq.bilayer-hamiltonian}, the latter will take the following  form
\begin{equation}
{\bf H}= \frac{1}{2m^*}
\left(\begin{array}{cc}
0 & \left( \left(p_{x}-\dfrac{eB_{0}}{2c} y\right) - i \left(p_{y}+\dfrac{eB_{0}}{2c} x \right)\right)^2 \\ 
\left( \left(p_{x}-\dfrac{eB_{0}}{2c} y\right) + i \left( p_{y}+\dfrac{eB_{0}}{2c} x \right)\right)^2 & 0
\end{array}
\right). 
\label{eq.bilayer-hamiltoian2}
\end{equation} 
It is now very convenient to introduce the following scalar operators, which naturally appear in \eqref{eq.bilayer-hamiltoian2},
\begin{equation} 
\mathcal{A}^{\pm}:=\sqrt{\dfrac{c}{2\hbar eB_{0}}}\left[ \left(p_{x}-\dfrac{eB_{0}}{2c} y \right)\pm i\left( p_{y}+\dfrac{eB_{0}}{2c} x    \right)  \right], \label{eq.apm}
\end{equation}
and which satisfy the following commutation relations
\begin{equation}\label{apm}
[\mathcal{A}^{-},\mathcal{A}^{+}]=\mathcal{I},\qquad
\mathcal{A}^{-}=( \mathcal{A}^{+})^{\dagger} ,
\end{equation}
where $\mathcal{I}$ is the identity operator. Therefore, we can say that the set $\{\mathcal{A}^{+},\mathcal{A}^{-},\mathcal{I}\}$ constitutes a Heisenberg (or boson) algebra.

Since the system has a geometric rotational symmetry about the $z$-axis, 
 it is more convenient to work in polar coordinates. Thus,  after a change to the normalized polar coordinates $(\xi,\theta)$ defined by
\begin{equation} 
x= r\cos \theta,\quad y = r \sin \theta,\qquad \xi:=\sqrt{\dfrac{{m^*} \omega_{c}}{2 \hbar}}\,r,\quad\ \omega_{c}=\frac{e B_{0}}{m^{*} c}, 
\end{equation}
where $\omega_{c}$ is the cyclotron frequency for non-relativistic electrons with effective mass  $m^{*}$, the operators $\mathcal{A}^{\pm}$  in \eqref{eq.apm} are rewritten as
\begin{equation}
\mathcal{A}^{\pm}
=\dfrac{\pm {i}\exp(\pm {i}\theta)}{2}\left[  \mp \partial_\xi-\dfrac{{i}}{\xi}\,\partial_\theta+ \xi  \right] .
\label{eq.op-A-polare}
\end{equation}
Therefore, the time-independent Dirac-type equation for ${\bf H}$ in \eqref{eq.bilayer-hamiltoian2} becomes
\begin{equation}
{\bf H}\,\Psi(\xi,\theta)=\hbar \omega_{c}
\left(\begin{array}{cc}
0 & ( \mathcal{A}^{-}) ^2 \\ 
( \mathcal{A}^{+}) ^2 & 0
\end{array}
\right)\Psi(\xi,\theta)=E\Psi(\xi,\theta).
\label{eq.Sch} 
\end{equation}
These eigenstates are two-component spinors 
\begin{equation}
\Psi(\xi,\theta)=
\left(\begin{array}{c}
\psi^{a}(\xi,\theta) \\  
\psi^{b}(\xi,\theta) 
\end{array} 
\right),
\label{eq.spinor}
\end{equation}   
which when taken to \eqref{eq.Sch}, generate the two coupled scalar equations
\begin{eqnarray}\label{eq.sol1111}
(\mathcal{A}^{+})^2 \psi^{a}(\xi,\theta)=\epsilon \, \psi^{b} (\xi,\theta), 
\\[1ex]
(\mathcal{A}^{-})^2 \psi^{b} (\xi,\theta)=\epsilon \,\psi^{a}(\xi,\theta), \label{eq.sol1}
\end{eqnarray}
where $\epsilon:= E/\hbar \omega_{c}$.
To decouple the previous equations, we square the Hamiltonian (\ref{eq.Sch}), obtaining the following two eigenvalue scalar equations
\begin{eqnarray}
\mathcal{H}_{1} \psi^{a}(\xi,\theta)=\epsilon^{2} \, \psi^{a}(\xi,\theta), 
& \quad &
\mathcal{H}_{1}:= (\mathcal{A}^{-})^2 (\mathcal{A}^{+})^2,
 \label{eq.sol2-1}
\\[1ex]
\mathcal{H}_{2} \psi^{b}(\xi,\theta)=\epsilon^{2} \,\psi^{b}(\xi,\theta), 
& \quad &
\mathcal{H}_{2}:= (\mathcal{A}^{+})^2 (\mathcal{A}^{-})^2.
 \label{eq.sol2-2}
\end{eqnarray}
In the next section we will address the explicit resolution of these two equations that
will allow us to find the spectrum of the bilayer Hamiltonian \eqref{eq.Sch}.
Note that the equations \eqref{eq.sol1111}--\eqref{eq.sol2-2} provide an example of supersymmetric partner Hamiltonians $\mathcal{H}_{1}$ and $\mathcal{H}_{2}$, connected  by second order operators $(\mathcal{A}^{\pm})^2$.

\section{Symmetries and spectrum of the scalar Hamiltonians $\mathcal{H}_{1,2}$}\label{simetrias}

To determine the spectrum of each of the two scalar Hamiltonians $\mathcal{H}_{1}$ and $\mathcal{H}_{2}$, we first find their symmetries and then use them to find the spectra.

\subsection{Basic symmetries of  $\mathcal{H}_{1}$ and $\mathcal{H}_{2}$}

Taking into account the rotational symmetry of the problem, as well as the definitions \eqref{eq.sol2-1}--\eqref{eq.sol2-2} of the two scalar Hamiltonians, we define the following auxiliary operators
\begin{equation}\label{eq.n-lz-operators}
\mathcal{L}_{z}:=-i\, \partial_\theta, \qquad
\mathcal{N}:=\mathcal{A}^{+}\mathcal{A}^{-}, 
\end{equation}
corresponding to the angular momentum operator along the direction $z$  and to the number operator of the boson algebra $\{\mathcal{A}^{+},\mathcal{A}^{-}, \mathcal{I}\}$, respectively. It is easy to check that the following commutation relations of all these operators hold
\begin{eqnarray}
\left[ \mathcal{L}_{z}, \mathcal{A}^{\pm} \right]= \pm\mathcal{A}^{\pm}, \qquad
\left[ \mathcal{N}, \mathcal{A}^{\pm} \right]= \pm\mathcal{A}^{\pm}, \qquad 
\left[ \mathcal{L}_{z}, \mathcal{N} \right]= 0.
\end{eqnarray}
Furthermore, using $[\mathcal{A}^{-},\mathcal{A}^{+}]=\mathcal{I}$, it is possible to express the scalar Hamiltonians $\mathcal{H}_{1}$ and $\mathcal{H}_{2}$ in terms of the number operator $\mathcal{N}$ as follows 
\begin{eqnarray}
\mathcal{H}_{1}=(\mathcal{N}+2\mathcal{I}) (\mathcal{N}+\mathcal{I}),\qquad
\mathcal{H}_{2}=\mathcal{N}(\mathcal{N}-\mathcal{I}).
\label{eq.hn}
\end{eqnarray}
Therefore, the operators $\mathcal{L}_{z}$ and $\mathcal{N}$ defined in \eqref{eq.n-lz-operators} are symmetries of
$\mathcal{H}_{1,2}$, that is, they satisfy the following commutation relations
\begin{eqnarray}
\left[ \mathcal{N},\mathcal{H}_{1,2}    \right]=0, \qquad  \left[ \mathcal{L}_{z} , \mathcal{H}_{1,2}  \right]=0.
\end{eqnarray}
Since the operators $\mathcal{N}$, $\mathcal{L}_z$,  and $\mathcal{H}_{1,2}$ commute, it is possible to search for simultaneous  eigenfunctions of all of them.
So, we could say that these states have an excitation number $n$ of bosons determined by the eigenvalues of $\mathcal{N}$ and a well-defined angular momentum  given by the eigenvalues of $\mathcal{L}_{z}$. If we create or annihilate an excitation boson through $\mathcal{A}^{\pm}$, then the state will also have a unit of angular momentum more or less.

We will start by finding a ground state of $\mathcal{N}$, such  that it is annihilated by the operator $\mathcal{A}^{-}$ and at the same time must also be an eigenstate of $\mathcal{L}_z$ (with eigenvalue $-m$, indicated this way for reasons that will be clarified later). That is, the ground state denoted by $\psi_{m,0} (\xi, \theta)$ will be characterized by
\begin{equation}
\mathcal{L}_z\psi_{m,0} (\xi, \theta) = -m \psi_{m,0} (\xi, \theta),\qquad
\mathcal{A}^{-} \psi_{m,0}(\xi, \theta)=0 .
\label{Pi-zero}
\end{equation}
Note that this is a pair of separable  differential equations in the variables 
$\theta$ and $\xi$ which can be solved to get their explicit solutions:
\begin{equation}
\psi_{m,0}(\xi, \theta)= C_{m,0} \, \xi^{m} \exp \left( - \xi^{2}/2-im \theta \right), \quad m=0,1,2,\ldots,  \label{ground-state}
\end{equation}
where $C_{m,0}$ are the corresponding normalization constants. Note that the possible values of $m$ have been restricted since only $m=0,1,2,\dots$ provides single-valued wave functions and gives rise to square integrable solutions. Now applying the creation operator $\mathcal{A}^{+}$ on \eqref{eq.op-A-polare} in each of the possible ground states, 
$\psi_{m,0}(\xi, \theta)$, it is possible to obtain the wave functions of the excited states, which after normalization turn out to be
\begin{equation}
\psi_{m,n}(\xi, \theta)= \dfrac{1}{\sqrt{n}!}\, (\mathcal{A}^{+})^{n} \, \psi_{m,0}(\xi, \theta), \qquad n=0,1,2,\ldots \label{excited states}
\end{equation}
It is not difficult to check that these functions satisfy the following eigenvalue equations
\begin{equation}
\mathcal{N}\psi_{m,n}(\xi, \theta)=n\,\psi_{m,n}(\xi, \theta),\qquad
\mathcal{L}_z\psi_{m,n}(\xi, \theta)=(n-m)\,\psi_{m,n}(\xi, \theta),
\end{equation}
and therefore have a well-defined value of angular momentum $\ell_z=n-m$. After some calculations, it is found that these normalized eigenfunctions $\psi_{m,n}(\xi, \theta)$, with $ m,n=0,1,2 \ldots$, turn out to be
\begin{equation}\label{psimnnormalizados}
\psi_{m,n}(\xi, \theta)=\sqrt{\dfrac{m^{*} \omega_{c}}{2 \pi \hbar}\, \dfrac{\min(m,n)!}{\max(m,n)!}}\,i^{n+m}\, (-1)^{\min(m,n)}  \exp\left[-\dfrac{\xi^{2}}{2}-i(m-n)\theta \right] \xi^{\vert m-n\vert} \, L^{\vert m-n \vert}_{\min(m,n)}(\xi^{2}),
\end{equation}
where $L^{\alpha}_{k}(x)$ are the associated Laguerre polynomials. 
Given \eqref{eq.hn}, the functions $\psi_{m,n}(\xi, \theta)$ will also be eigenfunctions of the scalar Hamiltonians $\mathcal{H}_{1,2}$, more precisely,
\begin{eqnarray}
\mathcal{H}_{1}\psi_{m,n}=(n+2)(n+1)\psi_{m,n}, \qquad
\mathcal{H}_{2}\psi_{m,n}= n(n-1) \psi_{m,n},\qquad m,n=0,1,2,\dots, \label{hn2}
\end{eqnarray}
and therefore these eigenfunctions $\psi_{m,n}$ can play the role of both the upper component 
$\psi^{a}$ and the lower component $\psi^{ b}$ of the spinor $\Psi(\xi,\theta)$ in \eqref{eq.spinor}.

\subsection{Additional symmetries of  $\mathcal{H}_{1}$ and $\mathcal{H}_{2}$}

As shown in the previous section, the discrete  spectrum of ${\cal N}$ is degenerate, and this degeneracy can be labeled with the index $m$, that is, all eigenfunctions $\psi_{m,n}$, $m =0, 1,2,\dots$, have the same eigenvalue $n$ of $\mathcal{N}$. This is a clear indication that there must be some additional symmetries responsible for  such degeneracy. To determine them, we introduce two differential operators on the variables $(\xi, \theta)$ $\mathcal{B}^{\pm}$, defined as follows
\begin{equation}
\mathcal{B}^{\pm}:=\dfrac{\pm {i}\exp(\mp {i}\theta)}{2}\left[ \mp\,\partial_\xi+\dfrac{{i}}{\xi}\,\partial_\theta+\xi \right], 
\label{eq.op-B-polares}
\end{equation}
which, together with the operators $\mathcal{A}^{\pm}$, serve to close the algebra 
\begin{equation}[\mathcal{B}^{-},\mathcal{B}^{+}]=\mathcal{I},
\qquad
\left[ \mathcal{A}^{\pm}, \mathcal{B}^{\pm} \right]=\left[ \mathcal{A}^{\pm}, \mathcal{B}^{\mp} \right]=0.
\end{equation}
Furthermore, the operators $\mathcal{B}^{\pm}$ satisfy the following commutation relations
with the symmetries $\mathcal{L}_{z}$ and  $\mathcal{N}$:
\begin{eqnarray}
\left[ \mathcal{L}_{z}, \mathcal{B}^{\pm} \right]= \mp\mathcal{B}^{\pm}, \qquad
\left[ \mathcal{N}, \mathcal{B}^{\pm} \right]=0.
\end{eqnarray}
Thus, summarizing, two Heisenberg commutative algebras have been obtained, generated by $\mathcal{A}^{\pm}$ and $\mathcal{B}^{\pm}$ and related by an inversion in the variable $\theta$, that is:
$\mathcal{B}^{\pm}=\mathcal{P} \mathcal{A}^{\pm}\mathcal{P}$, where 
$\mathcal{P}\psi (\xi, \theta)=\psi (\xi, -\theta)$.

In the same way that $\mathcal{N}$ has been defined associated with the operators $\mathcal{A}^{\pm}$, we can introduce another number operator $\mathcal{M}$ associated with the operators $\mathcal{B}^{\pm}$, which 
behaves like this
\begin{equation}\label{ms}
\mathcal{M}:=\mathcal{B}^{+}\mathcal{B}^{-} \quad \Longrightarrow\quad 
\left[ \mathcal{M}, \mathcal{B}^{\pm} \right]= \pm\mathcal{B}^{\pm},
\quad
\left[ \mathcal{M}, \mathcal{H}_{1,2} \right]= 0.
\end{equation}
Thus, we have arrived at three operators $\mathcal{L}_{z},\mathcal{M},\mathcal{N}$, which commute with each other and with $\mathcal{H}_{1,2}$, although they are not independent, because
\begin{equation}
\mathcal{N}-\mathcal{M}=-i \partial_\theta=\mathcal{L}_{z}.
\end{equation}
Note that the eigenfunctions $\psi_{m,n}$ derived above are labeled according to the eigenvalues of these operators
\begin{equation} \label{involutive}
\mathcal{N}\psi_{m,n}=n\psi_{m,n},\qquad  \mathcal{M}\psi_{m,n}=m\psi_{m,n}, \qquad \mathcal{L}_{z}\psi_{m,n}=(n-m)\psi_{m,n},\qquad m,n=0,1,2,\dots
\end{equation}
All these eigenfunctions $\psi_{m,n}$ are obtained by the successive action of the appropriate creation operators on what can be considered the genuine ground state, $\psi_{0,0}$, namely
\begin{equation}\label{psimn}
(\mathcal{B}^{+})^m(\mathcal{A}^{+})^n\psi_{0,0}=\sqrt{m!\, n!}\, \psi_{m,n},
\end{equation}
while
\begin{equation}
\mathcal{B}^{-}\psi_{0,0}=\mathcal{A}^{-}\psi_{0,0}=0.
\end{equation}

Summarizing, the symmetries of the scalar Hamiltonians $\mathcal{H}_{1}$ and $\mathcal{H}_{2}$ allow us to characterize all the eigenfunctions of the scalar problem associated with the bilayer Hamiltonian. In this sense we could say that $\mathcal{H}_{1}$ and $\mathcal{H}_{2}$ give superintegrable two-dimensional  systems with three independent symmetries. 
Next, we are going to use the results we have just found so far to study the properties of the bilayer Hamiltonian.

\section{Spectrum and properties of bilayer Hamiltonian ${\bf H}$}\label{spectrumbilayer}

Our next step is to find the symmetries of the bilayer Hamiltonian \eqref{eq.Sch}.
In this case it is convenient to write it as follows
\begin{equation}
{\bf H} =\hbar \omega_{c}
\left( (\mathcal{A}^{-}) ^2\,\sigma^+ + (\mathcal{A}^{+}) ^2\,\sigma^-\right),
\label{Scheq2} 
\end{equation}
where $\sigma^\pm =(\sigma_x\pm i\sigma_y)/2$, and $\sigma_i$, $i=x,y,z$, are the Pauli matrices. Note that (\ref{Scheq2}) resembles the Hamiltonian of the Jaynes-Cummings two-photon model \cite{Wodkiewicz,Gerry,Gerrybook}.

The scalar symmetries of Hamiltonians $\mathcal{H}_{1,2}$ given in (\ref{involutive}) must be adapted to become symmetries of the matrix Hamiltonian ${\bf H}$. If we denote by ${\bf I}$ the $2\times 2$ identity matrix, the symmetries of (\ref{Scheq2}) are implemented by the following operators
\begin{equation}\label{symm}
 {\bf B}^\pm=\mathcal{B}^\pm \,{\bf I}, 
 \quad
 {\bf M}= \mathcal{M}\,{\bf I},
\quad 
{\bf Q}= (\mathcal{N}+1) \,{\bf I}+ \sigma_z +1,
\quad 
{\bf J}_z = \mathcal{L}_z\, {\bf I} + \sigma_z= (\mathcal{N}-\mathcal{M}) \,{\bf I}+ \sigma_z,
\end{equation}
of which ${\bf M}$, ${\bf Q}$ and ${\bf J}_z$ commute with each other and are Hermitian, though not mutually independent, because they satisfy the obvious identity ${\bf J}_z= {\bf Q}- {\bf M}-{\bf I}$.
Therefore, the eigenspinors $\Psi_{m,q}$ can be characterized by three real numbers $m, q, j$ (of which only two of them are independent), such that the following holds
\begin{equation}\label{bisimetrias}
{\bf M}\Psi_{m,q}=m\Psi_{m,q},
\quad 
{\bf Q}\Psi_{m,q}={ q}\, \Psi_{m,q},
\quad 
{\bf J}_z\Psi_{m,q}={ (q-m-1)}\,\Psi_{m,q}.
\end{equation}
Each of these three operators has its own meaning:

\begin{itemize}
\item
${\bf M}$ is the scalar operator introduced previously in \eqref{ms}. 
This is the number operator for the symmetries $\mathcal{B}^\pm$ spanning each degeneracy eigenspace. These symmetries are related to displacements in a constant magnetic field that maintains the energy \cite{Manko}.

\item
${\bf Q}$ is a kind of excitation number and it is not scalar (in the sense that it does not commute with pseudo-spin operators $\sigma_z,\sigma^\pm$). 
This is specific of our bilayer Hamiltonian, and tell us that the excitation of a state depends on the Landau level (via 
$\mathcal{N}$)  as well as on the type of the bilayer atom (via $\sigma_z$). In its definition in (\ref{symm}), we have added a unit to have only positive eigenvalues: $q=0,1,2,\dots$ Note that although $\mathcal{N}$ is a symmetry in the scalar Hamiltonians, this is no longer the case in bilayer graphene, where it could be replaced by ${\bf Q}$. 

\item
${\bf J}_z$ is similar  to the total angular momentum in the $z$ axis. 
The term $\mathcal{L}_z$ is the orbital part, while $\sigma_z$ is the pseudo-spin part $s_z=1$, $\sigma_z=1\oplus(-1)$. Therefore ${\bf J}_z$ is not a scalar because it has a different action on each  component of the spinor. The operator $\sigma_z$ has two eigenvalues $\pm 1$ that can be interpreted as two helicity  values of pseudo-spin $s_z=1$ \cite{mccannsolo}. Formally, the total angular momentum is a tensor sum and the states (the spinors) are  products of orbital and pseudo-spin spaces,
\[
{\bf J}_z=\mathcal{L}_z\oplus \sigma_z,\qquad j_z = \ell\otimes (1\oplus(-1))= (\ell+1)\oplus(\ell-1).
\]
Note that this symmetry is different from the one-layer graphene where ${\bf J}_z = \mathcal{L}_z\, {\bf I} + \frac12\sigma_z$, corresponding to pseudo-spin $s_z=1/2$. Due to this property, there are important consequences that are different in monolayer and bilayer graphene. For example, a bilayer spinor is invariant under the rotation of $2\pi$  around the $z$-axis, while the monolayer will change sign, Berry's phase will be different under a full rotation and the scattering properties will be different too \cite{MF2006}.
\end{itemize}

Consider next the eigenspaces $V_{m,q}$ and the eigenspinors $\Psi_{m,q}$ of the  symmetries 
${\bf M}$ and ${\bf Q}$, taking into account that their discrete eigenvalues are $m=0,1,2\dots$ and $q = 0,1,\dots$, respectively. They can be expressed in terms of the scalar functions $\psi_{m,n}$ \eqref{psimnnormalizados} as follows:
\begin{equation}
\begin{array}{ll}
V_{m,0}=\left\langle \Psi_{m,0} =\left(\begin{array}{c}
0\\ 
\psi_{m,0}\end{array}\right)  \right\rangle\,,
\quad 
V_{m,1}=\left\langle  \Psi_{m,1} =\left(\begin{array}{c}
0\\ 
\psi_{m,1}\end{array}\right) 
\right\rangle\,,\quad &q=0,1
\\ [3.5ex]
V_{m,q}=\left\langle  \Psi^{\uparrow}_{m,q} = \left(\begin{array}{c}
\psi_{m,q-2}\\[0.75ex]
0\end{array}\right)\,, \quad 
 \Psi^{\downarrow}_{m,q} =\left(\begin{array}{c}
0\\ 
\psi_{m,q}\end{array}\right)
\right\rangle\,,\qquad &q\geq 2.
\end{array}
\end{equation}
The first two eigenspaces, $V_{m,0}$ and $V_{m,1}$, are one-dimensional, and the others, $V_{m,q}$, $q\geq 2$, are two-dimensional.
Next, the eigenvalues and  eigenstates of the Hamiltonian in \eqref{Scheq2} will be calculated by restricting ${\bf H}$ to each of these eigenspaces. Note that the eigenvalues can be positive or negative, hence the notation $\pm$ in the following expressions (remark that from now on, we will replace the eigenvalue subscript $q$  in $\Psi_{m,q}$ by the $n$-value of the lower component of the spinor, that is: changing $\Psi_{m,q}$ by $\Psi_{m,n}$, with $n = q$, which is a slight abuse of notation).

For $n\geq 2$ (or $q\geq 2$), we will write the solutions in the form
\begin{equation}
\Psi_{m,n}^\pm(\xi, \theta):=\dfrac{1}{\sqrt{2}}
\left(\begin{array}{c}
\psi_{m,n-2}(\xi, \theta) 
\\[1ex] 
\pm\,\psi_{m,n}(\xi, \theta) 
\end{array}
\right),  \quad 
E_n^\pm=\pm \hbar \omega_{c} \sqrt{n(n-1)}, \quad n=2,3,\ldots,
\label{eq.eigenstates}
\end{equation} 
and for the fundamental zero energy solutions  $n=0,1$ (or $q=0,1$),   we will have the expressions
\begin{equation}
\Psi_{m,0}^\pm(\xi, \theta):=
\left(\begin{array}{c}
0 
\\ 
\pm\,\psi_{m,0}(\xi, \theta) 
\end{array}
\right), \ E_0^\pm=0,\qquad 
\Psi_{m,1}^\pm(\xi, \theta):=\left(\begin{array}{c}
0 
\\ 
\pm\,\psi_{m,1}(\xi, \theta) 
\end{array}
\right),\ E_1^\pm=0.
\label{psi00}
\end{equation}
Some of the most important features of the solution we have just found are the following: (i) the spectrum is symmetric in positive and negative eigenvalues; (ii) each energy level is infinitely degenerate and within each subspace of equal energy the degeneracy is labeled with the index $m=0,1,2\dots$; (iii) in addition to this already mentioned degeneracy, the energy level $E=0$ has a double degeneracy for each value of $m$, as follows from \eqref{psi00}, and taking into account the non-equivalent $K'$ Dirac point and the true spin of the electron, the degeneracy for each eigenspace with the same value of $m$ will be 4 for $E_n\neq0$, and 8 for $E_0=E_1=0$; (iv) the eigenfunctions $\Psi_{m,n}^\pm(\xi, \theta)$ of ${\bf H}$ are also eigenfunctions of the pseudo-spin operator ${\bf J}_z$ with eigenvalue $j=n-m-1$.

After adequately elucidating the symmetries and the spectrum of the bilayer graphene Hamiltonian, our interest will be directed to the definition of probability and current densities, in polar coordinates. We will pay special attention to the current density, which depends strongly  on the Hamiltonian and it is not well known for the case
of the bilayer Hamiltonian (some results without magnetic field are given in \cite{FGN2011}).

\subsection{Probability densities for energy eigenstates}

The radial probability density $\rho(\xi,\theta)$ is defined in the usual way and in particular, for  eigenfunctions 
$\Psi_{m,n}^{\pm}$, it will be denoted by $\rho_{m,n}(\xi)$. It can be written for the present case in compact form as follows
\begin{equation}
\rho_{m,n}(\xi)=\left( \Psi_{m,n}^{\pm}(\xi, \theta)\right)^{\dagger} \Psi_{m,n}^{\pm}(\xi, \theta)=
\dfrac{ (1-\delta_{n0}-\delta_{n1})\vert \psi_{m,n-2}(\xi, \theta)\vert^{2}+\vert\psi_{m,n}(\xi, \theta)\vert^{2}}{2^{ (1-\delta_{n0}-\delta_{n1})}},\quad n=0,1,2,\dots
\label{eq.rhomn}
\end{equation}
\begin{figure}[htb]
  \begin{center}
    \subfigure[$m=0$, $n=2$]{
        \includegraphics[width=0.34\textwidth]{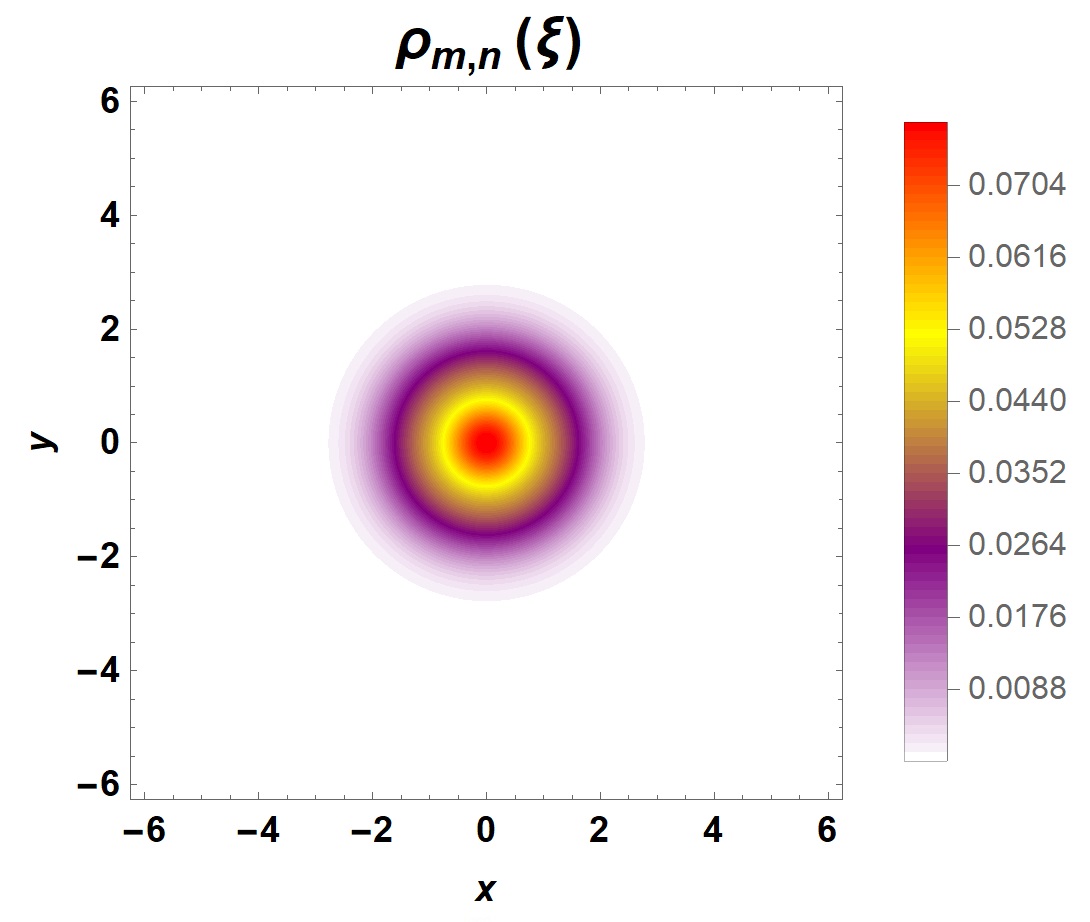} 
        \label{density m=01}}\qquad
    \subfigure[$m=1$, $n=2$]{
        \includegraphics[width=0.34\textwidth]{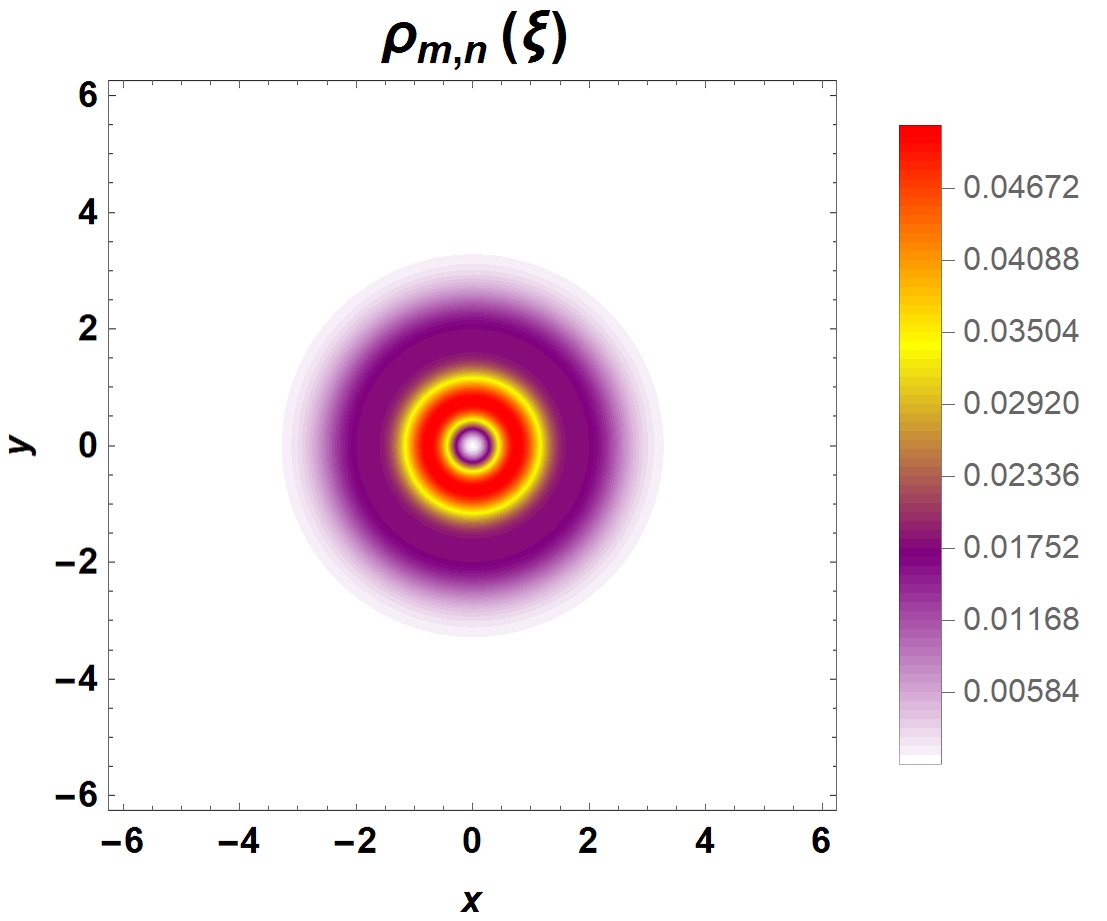} 
        \label{density-00}} 
        \subfigure[$m=2$, $n=2$]{
        \includegraphics[width=0.34\textwidth]{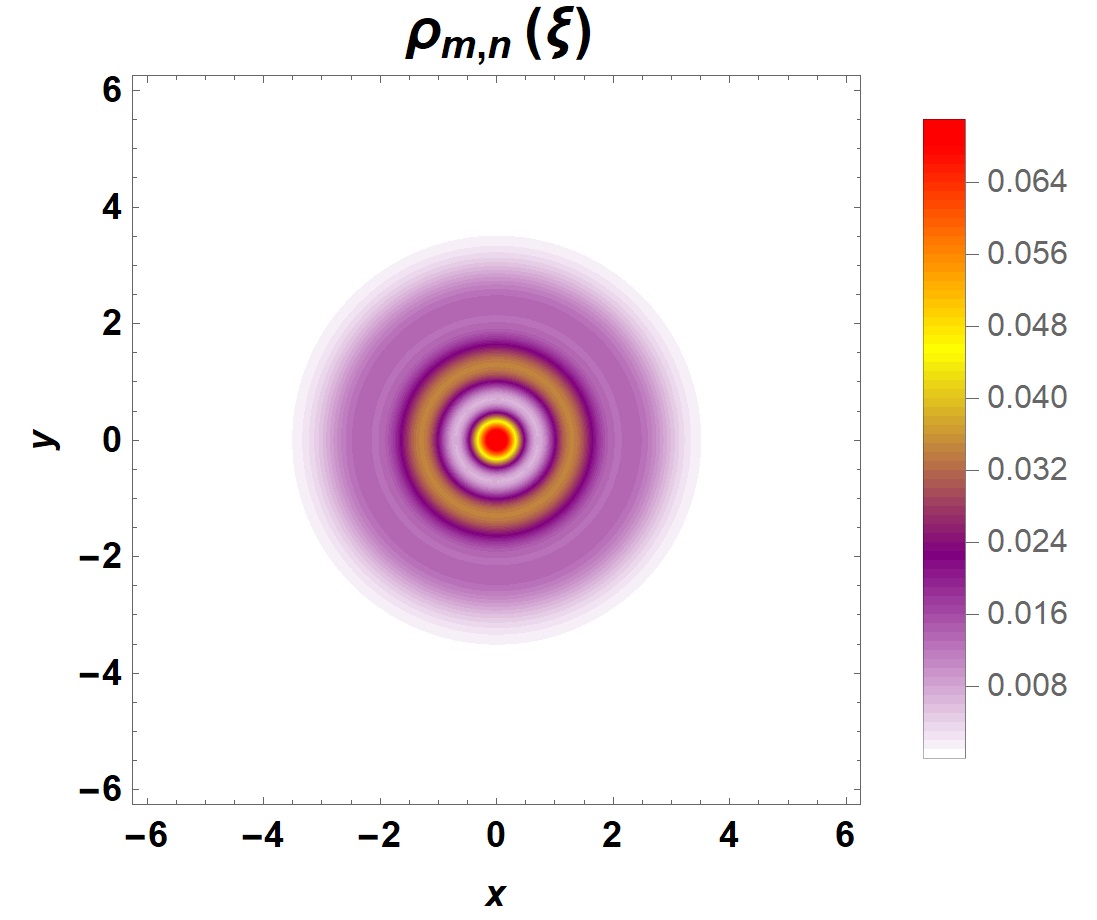} 
        \label{density-00}}\qquad
        \subfigure[$m=3$, $n=2$]{
        \includegraphics[width=0.34\textwidth]{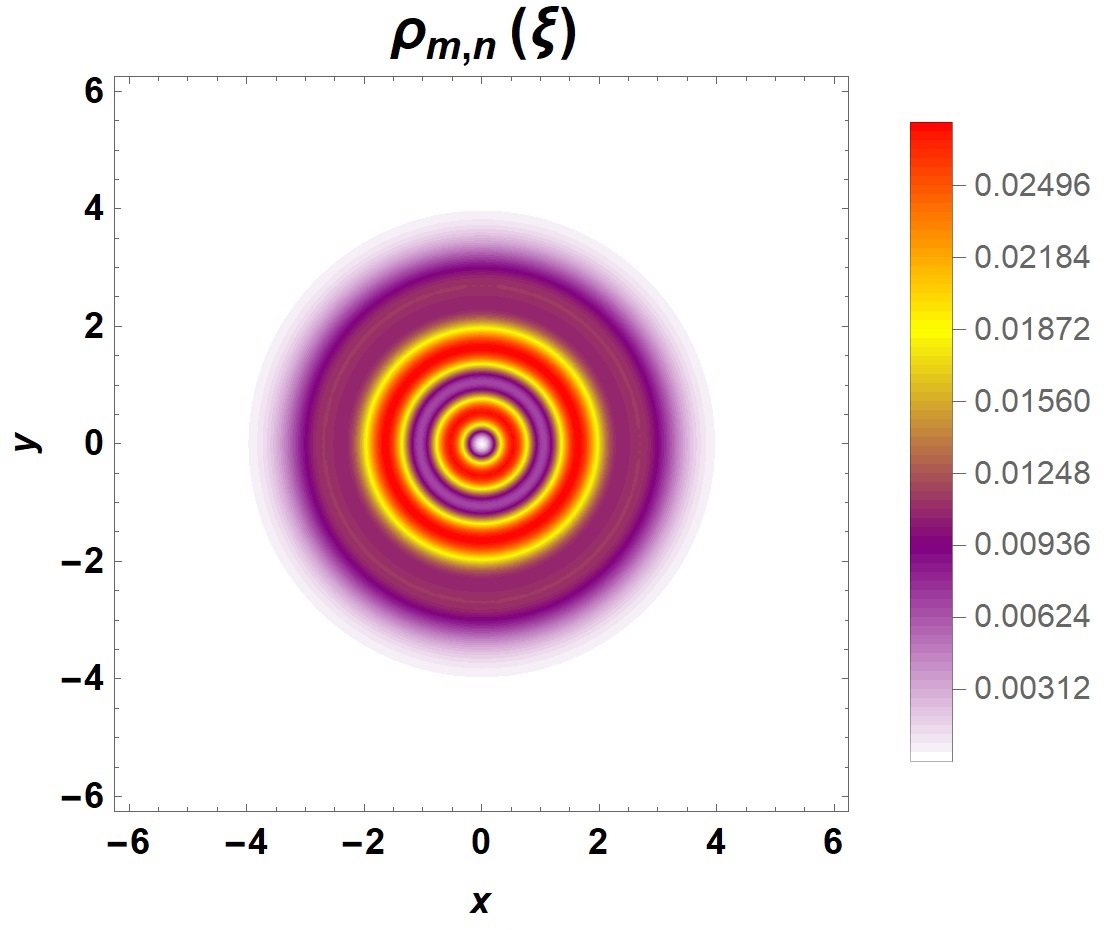} 
        \label{density-00}} 
    \caption{Probability densities for the eigenstates $\Psi^{\pm}_{m,n}(\xi, \theta)$ evaluated using \eqref{eq.rhomn} 
    for $n=2$ and various values of $m$. Graphs (a)--(d) show the corresponding densities, the colors indicate the different values.}
    \label{probability-densities-eigenstates}
  \end{center}
\end{figure}
Taking into account \eqref{psimnnormalizados}, it is clear that this expression does not depend on the angular variable $\theta$, and is the same for positive and negative energy eigenfunctions $\Psi_{m,n}^{\pm}$.
Figure~\ref{probability-densities-eigenstates} shows the probability density described in \eqref{eq.rhomn} for $n=2$ and several values of the index $m$. 
As $m$ increases, the centrifugal force forces the density away from the origin. However, when the value of the angular momentum $\ell_{z}=n-m$ is equal to or close to zero, the probability density for the energy eigenstates of bilayer graphene is mainly centered at the origin. The radial symmetry of the magnitude analyzed is evident in the plots.

\subsection{Current densities for energy eigenstates}

On the other hand, regarding the  current density for electrons in bilayer graphene, the appropriate expression for the free case ($B_0=0$) was derived in \cite{FGN2011} obtaining that the components $J_{k}$ of the current have the form
\begin{equation}
J_{k}[\Psi]=\dfrac{\hbar}{m^{*}}
\, \operatorname{Im} \left( \Psi^{\dagger}\, {\bf J}_{k}\, \Psi \right).
\label{eq.current}
\end{equation}
In the case of Cartesian coordinates, that is, with $k=x,y$, the current density operators $({\bf J}_{x},{\bf J}_{y})$ are given by 
\begin{equation}
{\bf J}_{x}=\dfrac{i}{\hbar}
\left(
\begin{array}{cc}
0 & p^-  \\ 
p^+& 0
\end{array}
\right), 
\qquad
{\bf J}_{y}=\dfrac{1}{\hbar}
\left(
\begin{array}{cc}
0 & p^-  \\ 
-p^+& 0
\end{array}
\right),\qquad p^\pm= p_x\pm i p_y.
\label{current-xy}
\end{equation}
It can be verified that $({\bf J}_{x},{\bf J}_{y})$ behaves like a vector under
the rotations ${\bf J}_{z}$ given in \eqref{bisimetrias}, so that the divergence  
$\partial_x{\bf J}_{x}+\partial_y{\bf J}_{y}$ is a scalar.

In the situation where there is a constant magnetic field perpendicular to the surface, $\vec{B}= B_{0}\vec{k}$, described by a potential in a symmetric gauge \eqref{eq.vector-potential}, the previous expressions for the current density operators are no longer valid for the bilayer Hamiltonian. However, after a somewhat long calculation we arrive at a continuity equation
\begin{equation}
\frac{\partial J_{x}}{\partial x}+ \frac{\partial J_{y}}{\partial y}+\frac{\partial \rho}{\partial t}=0, \label{eq.continuity}
\end{equation}
where the correct expression for ${\bf J}_{x},{\bf J}_{y}$ in (\ref{eq.current}) is
\begin{equation}
{\bf J}_{x}=\dfrac{i}{\hbar}
\left(
\begin{array}{cc}
0 & \pi^-  \\ 
\pi^+& 0
\end{array}
\right), 
\qquad
{\bf J}_{y}=\dfrac{1}{\hbar}
\left(
\begin{array}{cc}
0 & \pi^-  \\ 
-\pi^+& 0
\end{array}
\right),
\label{current-xyb}
\end{equation}
where  the momentum operators $p^\pm$ in \eqref{current-xy} for the free case must be replaced by
\begin{equation}
p^- \to \pi^-= p^- - \dfrac{eB_{0}}{2c}\left( y+ix\right) , 
\qquad 
p^+ \to \pi^+= p^+ - \dfrac{eB_{0}}{2c}(y-ix),
\end{equation}
according to the minimal coupling rule. Thus, the current density in bilayer graphene is gauge invariant because the derivatives have been replaced by covariant derivatives. 

On the other hand, in polar coordinates, the ones that concern us in the present work, the components $k=\xi, \theta$ of the current density operator are 
$({\bf J}_{\xi}, {\bf J}_{\theta})$. Remember that in polar coordinates, the divergence of a generic vector field, 
${\bf V}=({V}_{\xi}, {V}_{\theta})$, which must be placed in the continuity equation (\ref{eq.continuity}) is
\[
\nabla \cdot {\bf V}= 
\frac1{\xi} \frac{\partial }{\partial \xi} (\xi\,{V}_{\xi})
+\frac1{\xi} \frac{\partial }{\partial \theta}({V}_{\theta} ).
\]
Then,  the same procedure applied to the Cartesian coordinates in \eqref{eq.current}, after quite a long calculation leads to the components ${\bf J}_{k}$ of the current density in polar coordinates 
\begin{equation}
{\bf J}_{\xi}=
i
\left(
\begin{array}{cc}
0 & e^{-i\theta}\mathcal{A}^-  \\
e^{i\theta}\mathcal{A}^+ & 0
\end{array}
\right), \qquad 
{\bf J}_{\theta}=
\left(
\begin{array}{cc}
0 & e^{-i\theta}\mathcal{A}^-   \\ 
-e^{i\theta}\mathcal{A}^+& 0
\end{array}
\right), 
\label{current-rtheta}
\end{equation}
where $\mathcal{A}^{\pm}$ are the operators given in \eqref{eq.op-A-polare}.
It can be verified that both components are scalar, that is, they commute with ${\bf J}_{z}$ given in (\ref{symm}), and its divergence, in the continuity equation, is also another scalar.

\begin{figure}[htb]
  \begin{center}
    \subfigure[$m=0$, $n=2$]{
        \includegraphics[width=0.34\textwidth]{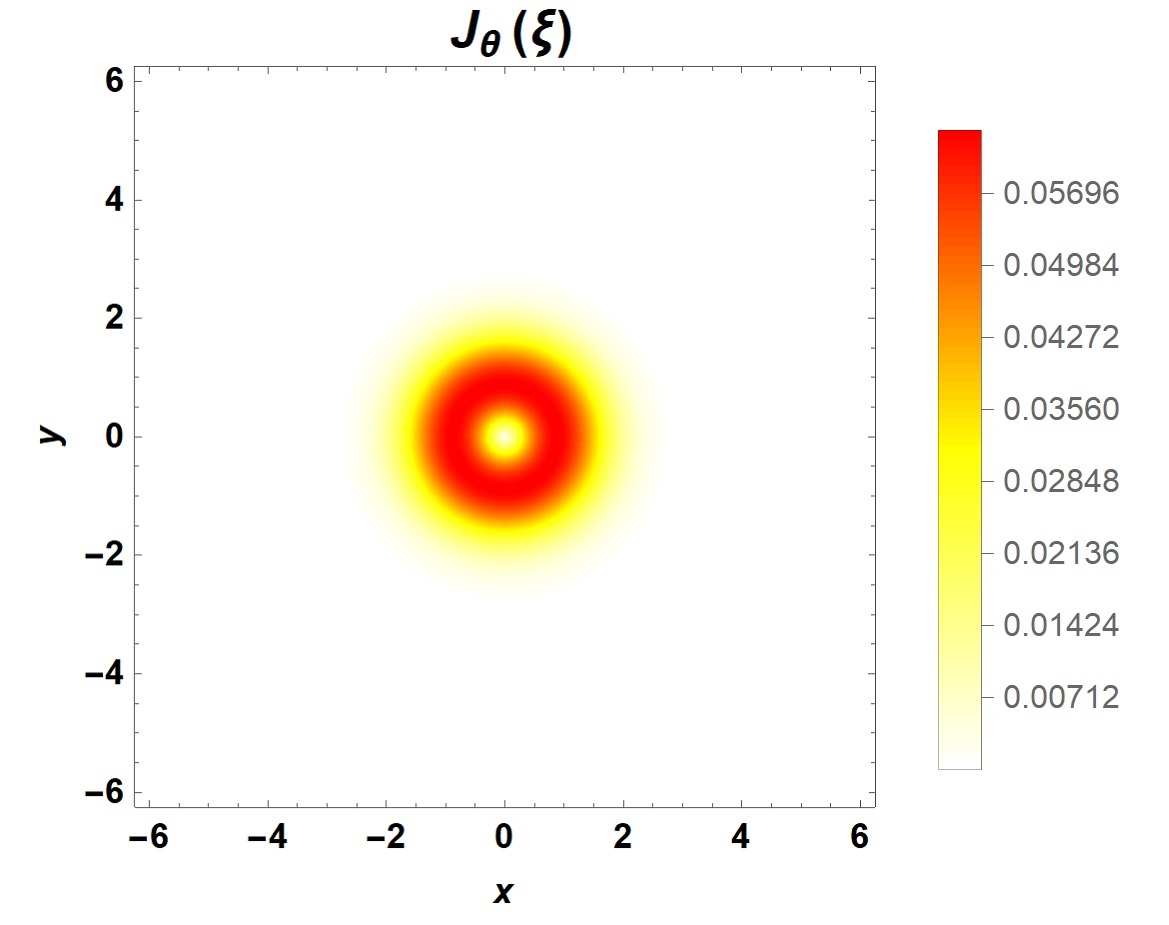}
        \label{density m=01}}\qquad
    \subfigure[$m=1$, $n=2$]{
        \includegraphics[width=0.34\textwidth]{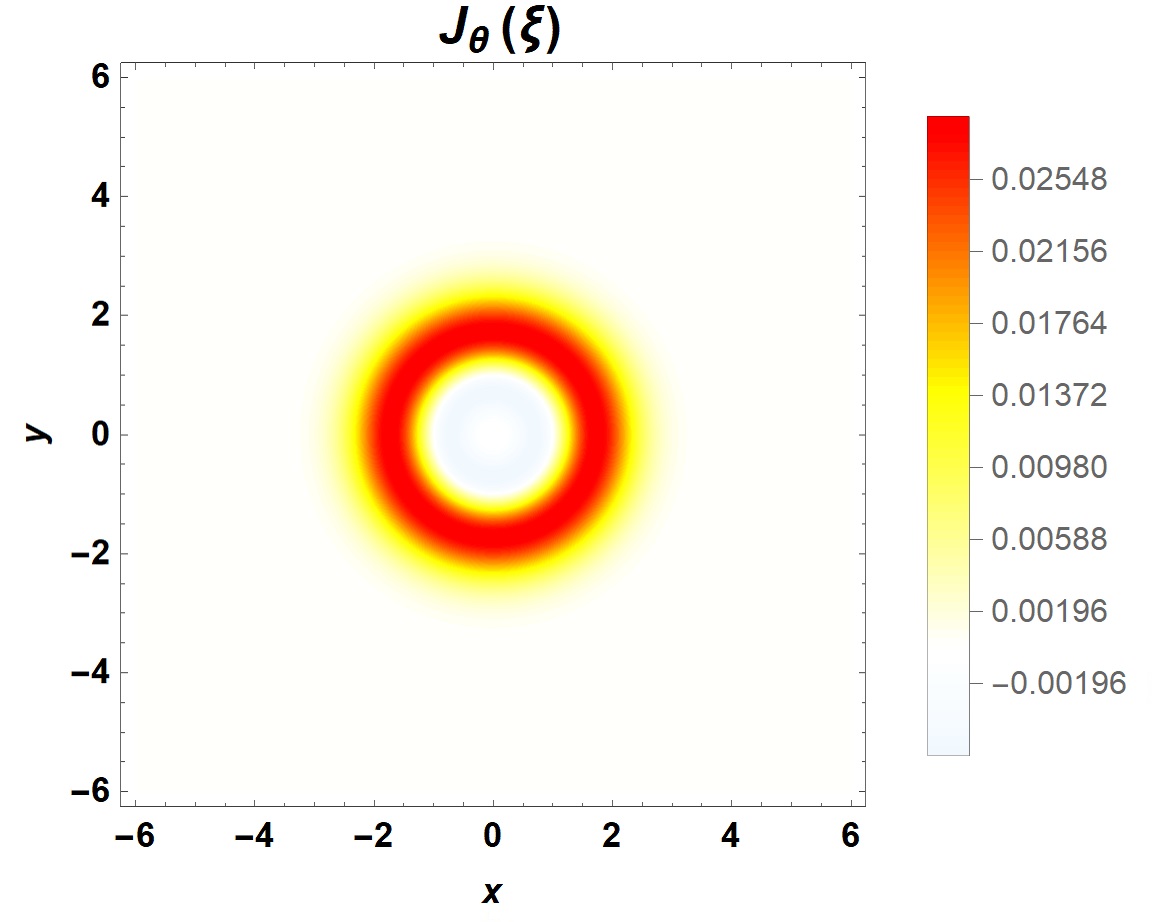}
        \label{density-00}}  
         \subfigure[$m=2$, $n=2$]{
        \includegraphics[width=0.34\textwidth]{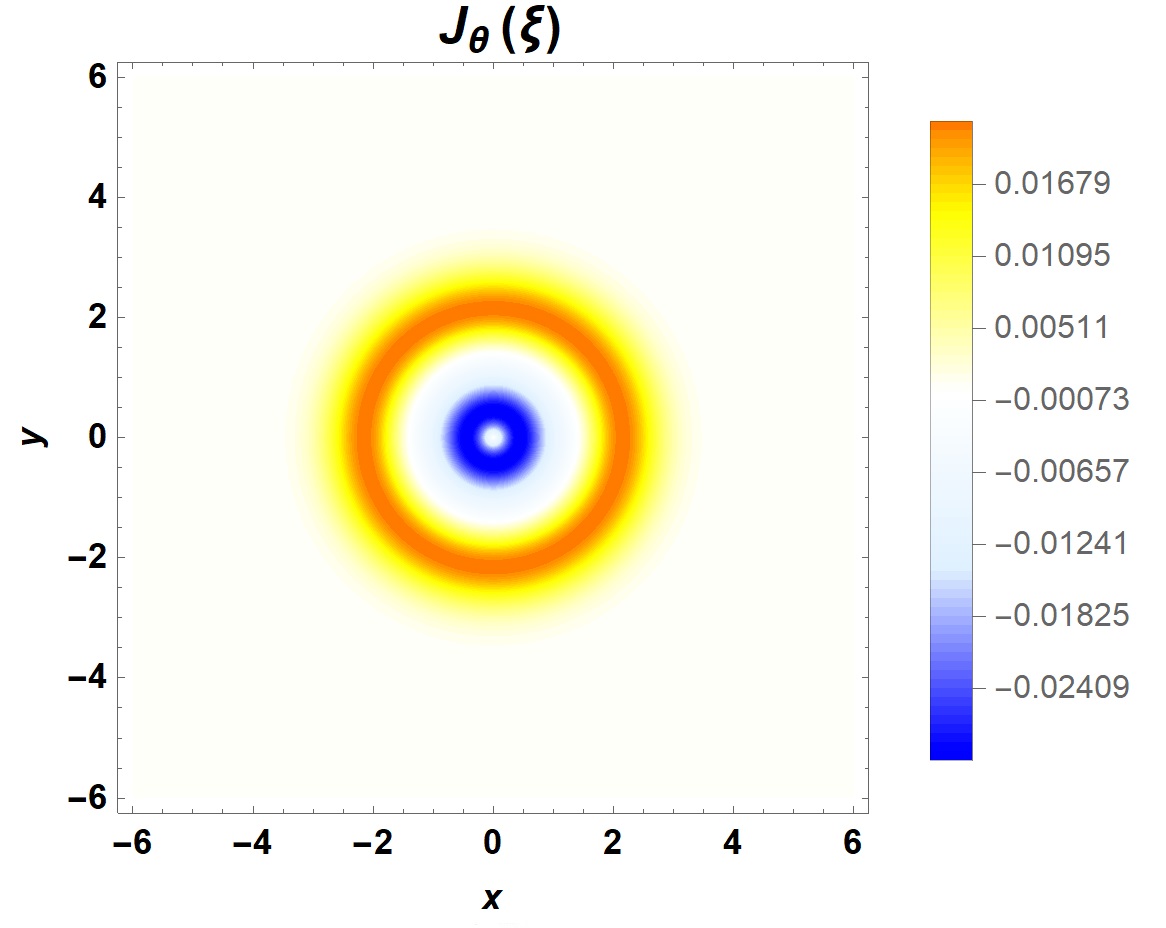}
        \label{density-00}}\qquad
         \subfigure[$m=3$, $n=2$]{
        \includegraphics[width=0.34\textwidth]{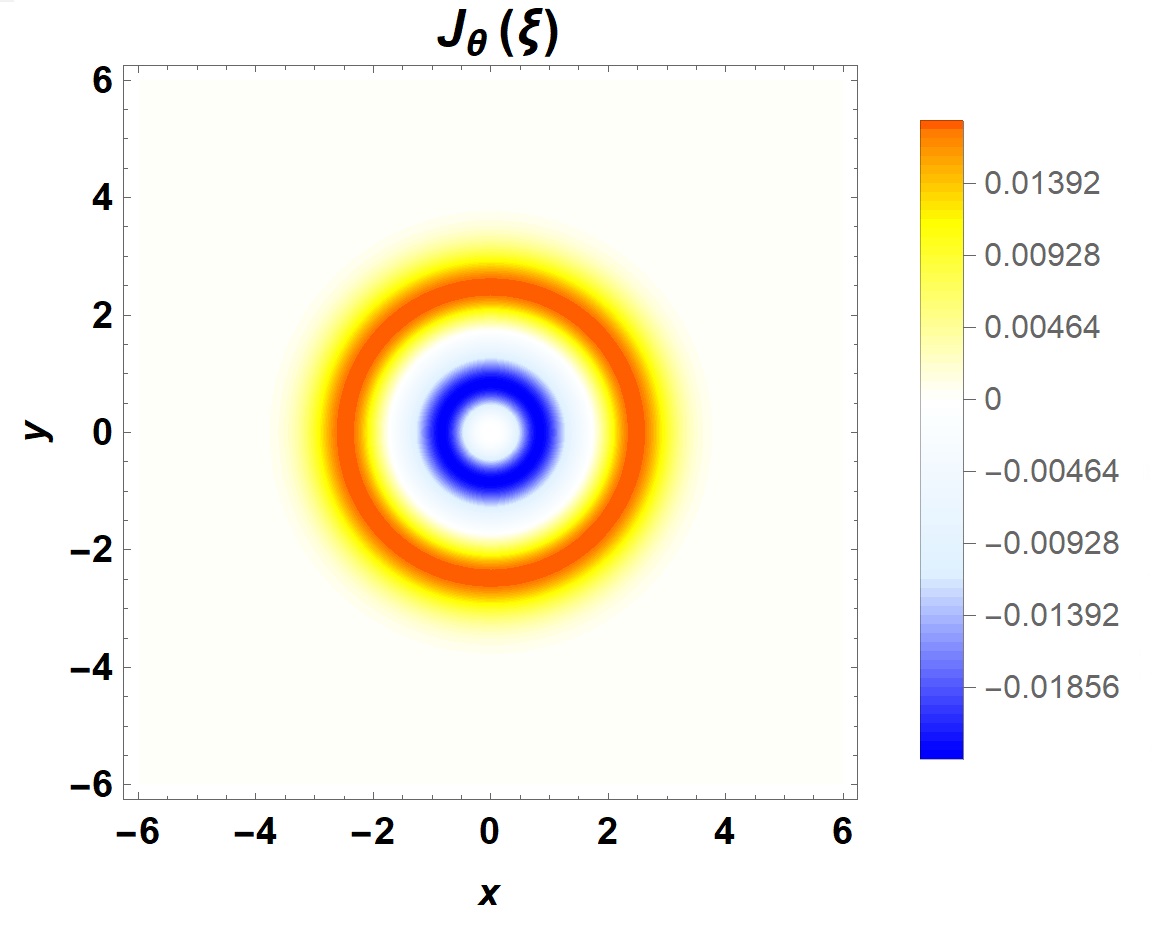}
        \label{density-00}} 
    \caption{Current density $J_{\theta}(\xi)$ for the eigenstates $\Psi^{\pm}_{m,n}(\xi, \theta)$ for $n=2$ and  $m=0,1,2,3$.}
    \label{current-theta-eigen}
  \end{center}
\end{figure}

In this way, the radial and tangential current densities for the energy states of bilayer graphene can be calculated by applying the operators \eqref{current-rtheta} to the states $\Psi^{\pm}_{m,n}(\xi, \theta)$.
The main result for these currents due to eigenstates is the following
\begin{equation}
J_{\xi}[\Psi^{\pm}_{m,n}]=0\,, 
\qquad 
J_{\theta}( \xi )[\Psi^{\pm}_{m,n}] \neq 0 \,,
\end{equation} 
for states with $n\geq2$. In other words, the radial current always disappears, while the tangential current remains and  only depends on the radius $\xi$ for each eigenspinor. 
To illustrate these results, Figure~\ref{current-theta-eigen} shows some graphs of its behavior for $n=2$ and various values of the $m$ index. 
Note that the results depend neither on the angular variable $\theta$ nor on the sign ($\pm$) of the energy. 
However, when the angular momentum $\ell_{z}<2$, then the angular probability current has a highly oscillating shape showing the existence of rings in the graphene bilayer in which the current density rotates clockwise or counterclockwise.

Once we have thoroughly familiarized ourselves with both the symmetries and the various physical properties of the eigenstates of the Hamiltonian describing bilayer graphene, in the next section we will go one step further, calculating 
the coherent states that can appear naturally in the bilayer Hamiltonian and discussing how the symmetry operators can help us interpret them.

\section{Coherent states of the  bilayer Hamiltonian ${\bf H}$}\label{coherentbilayer}

It does not seem necessary to insist on the great relevance of coherent states, both from the theoretical and experimental point of view, in various fields of Physics \cite{coherente,Pe86,AAG2014}.  
For this reason we are going to construct the coherent states of the  bilayer Hamiltonian ${\bf H}$ below, and for that in what follows we will adopt the standard approach, defining coherent states as eigenstates of some non-Hermitian annihilation operator that must be chosen appropriately according to the Hamiltonian.
Thus, the coherent states of the bilayer Hamiltonian $\bf H$, henceforth denoted as ${\Psi_\alpha(\xi,\theta)}$, will be defined as eigenstates of an annihilation operator
${\bf A}^{-}$, to be defined later, with complex eigenvalue depending on the complex number $\alpha$.
Since each energy eigenspinor $\Psi_{m,n}^{\pm}$ involves the scalar wave functions $\psi_{m,n}$ and $\psi_{m,n-2}$, it appears that the most economical way to define the operator ${\bf A}^{-}$ by its action on the eigenstates $\Psi_{m,n}^{\pm}(\xi,\theta)$ of \eqref {eq.eigenstates}--\eqref{psi00}, must have the form
\begin{equation}
{\bf A}^{-}\Psi_{m,n}^{\pm}(\xi,\theta)=g(n)\, \Psi_{m,n-2}^{\pm}(\xi,\theta), 
\label{eq.annihilation-operator}
\end{equation}
where $g(n)$ is a function to be determined. We will assume our coherent state will have a constant value of $m$ in (\ref{eq.annihilation-operator}), and therefore we will add the corresponding subscript to it:
\[
{\bf M}\, \Psi_{m,\alpha}= m\, \Psi_{m,\alpha}.
\]
Other options \cite{DNN19,DNN21} could have been chosen, such as coherent states with a fixed momentum 
${\bf J}_z$-eigenvalue or with a given value of energy, but our purpose here is to show, as an example, how to apply the symmetry operators in the context of bilayer coherent states.

Due to the fact that the bilayer Hamiltonian operator (\ref{eq.Sch}) has squared scalar creation and annihilation operators, $(\mathcal{A}^{\pm})^2$, which act on the scalar components of $\Psi_{m,n}^{\pm}$, the eigenvalue of ${\bf A}^{-}$ in \eqref{eq.coherent-states} is  taken to be $\alpha^2$, instead of the usual $\alpha$ value (in other words, this choice is suggested by the ``two-photon'' character of the bilayer Hamiltonian):
\begin{equation}
{\bf A}^{-}\, {\Psi_{m,\alpha}^{\pm}( 
 \xi,\theta)}=\alpha^{2}\, {\Psi_{m,\alpha}^{\pm}(
 \xi,\theta)}, \qquad \alpha \in\mathbb{C}, 
\label{eq.coherent-states}
\end{equation}
After these considerations, a coherent state will have the following general expression
\begin{equation}
\Psi_{m, \alpha}^\pm(  
 \xi,\theta)= \sum_{n=0}^{\infty} a_{n}\Psi_{m,n}^{\pm}(\xi,\theta),
\label{linear-combination}
\end{equation}
with certain complex coefficients $a_n$. Note that in \eqref{linear-combination} the coherent states also include a sign $\pm$ to distinguish between coherent states of positive or negative energy, built respectively from positive or negative energy eigenfunctions.
Taking this expression to \eqref{eq.coherent-states} and using \eqref{eq.annihilation-operator} we get
\begin{equation}
\sum_{n=0}^{\infty}\left(g(n+2)a_{n+2}-\alpha^{2} a_{2n} \right) \Psi_{m,n}^{+}(\xi,\theta)=0.
\end{equation} 
Given that the states $\Psi_{m,n}^{+}(\xi,\theta)\neq0$, $\forall n$, from the above equation we arrive at two independent recurrence relations for the coefficients $ a_{n}$ in terms of the free parameters $\left\lbrace a_{0}, a_{1} \right\rbrace$:
\begin{equation}
a_{2n}=\dfrac{\alpha^{2n}}{[g(2n)]!}a_{0}, \quad a_{2n+1}=\dfrac{\alpha^{2n}}{[g(2n+1)]!}a_{1}, \qquad n=0,1,2,\ldots,
\end{equation} 
where
\begin{equation}
[g(2n)]!:=\left\{
	       \begin{array}{cl}
		 1,  &\textrm {for} \  \ n=0, \\ [1ex]
		 g(2) \cdots g(2n), &\textrm{for} \ \ n>0,
		 \end{array}
	     \right.  \quad [g(2n+1)]!:=\left\{
	       \begin{array}{cl}
		 1,  &\textrm {for} \  \ n=0, \\ [1ex]
		 g(3) \cdots g(2n+1) , & \textrm{for} \ \ n>0.
		 \end{array}
	     \right.
	     \label{eq.function-g}
\end{equation}
Among the several possibilities to choose the function $g(n)$, there is one that is extremely natural, given the structure that these functions have in \eqref{eq.function-g}, it is $g(n) =\sqrt{n(n-1)}$. With this choice, the two sets of normalized coherent states that arise by taking only odd or even indices, denoted respectively ${\Psi_{m,\alpha}^{\it e \pm} ( \xi,\theta)}$ and ${\Psi_{m,\alpha}^{\it o \pm} ( \xi,\theta)}$, are these
\begin{eqnarray}
&&   {\Psi_{m,\alpha}^{e\pm} ( \xi,\theta)} = \dfrac{1}{\sqrt{\cosh(\vert \alpha \vert^{2})}}\left[ \Psi_{m,0}^{\pm}(\xi,\theta)+ \sum_{n=1}^{\infty}\dfrac{\alpha^{2n}}{\sqrt{(2n)!}}\Psi_{m,2n}^{\pm}(\xi,\theta)\right] , \label{coherenteven1}
\\ [1ex]
&& {\Psi_{m,\alpha}^{o\pm} ( \xi,\theta)} = \dfrac{1}{\sqrt{\sinh(\vert \alpha \vert^{2})/\vert \alpha \vert^{2}}}\left[ \Psi_{m,1}^{\pm}(\xi,\theta)+ \sum_{n=1}^{\infty}\dfrac{\alpha^{2n}}{\sqrt{(2n+1)!}}\Psi_{m,2n+1}^{\pm}(\xi,\theta)\right],
\label{coherentodd1}
\end{eqnarray}
where both the explicit form of the eigenstates \eqref{eq.eigenstates}--\eqref{psi00} and their normalization \eqref{psimnnormalizados} have already been taken into account.

\subsection{Expectation values}\label{expecatationvalues}

One of the important aspects of a coherent state is the interpretation of the complex parameter $\alpha$ on which it depends in terms of its position (and also its momentum, but this will not be discussed here).
Therefore, we start by finding the expectation value of $\xi^{2}$ in one of the coherent states derived above. To do this we start by expressing $\xi^{2}$ in terms of the scalar symmetry operators $\mathcal{A}^{\pm}$ and $\mathcal{B}^{\pm}$. From \eqref{eq.op-A-polare} and \eqref{eq.op-B-polares} we obtain
\begin{equation}
\xi^{2}=(\mathcal{A}^{+}{-}\mathcal{B}^{-})(\mathcal{A}^{-}{-}\mathcal{B}^{+}). 
\end{equation}
Therefore, it is possible to calculate the expectation value of $ \xi^{2} $ for instance in the even and positive energy coherent states \eqref{coherenteven1} as follows
\begin{equation}
\langle \xi^{2} \rangle_{m,\alpha}
=\langle \Psi_{m,\alpha}^{e\pm} \vert \, (\mathcal{A}^{+}{-}\mathcal{B}^{-})(\mathcal{A}^{-}{-}\mathcal{B}^{+})\, {\bf I}\, 
\vert \Psi_{m,\alpha}^{e\pm}\rangle.   
\end{equation}
Note that exactly the same can be done for odd coherent states.
Thus, after performing the integration over the spatial variables and simplifying, the expected value of $\xi^{2}$ turns out to be
\begin{equation}\label{xixi}
\langle \xi^{2}  \rangle^e_{m,\alpha} 
= \dfrac{1}{\cosh(\vert \alpha \vert^{2})}\left[\sum_{n=1}^{\infty} \dfrac{\vert \alpha \vert^{4n}}{(2n)!}(2n-1)\right]+(m+1) .  
\end{equation}
In particular, for $m=0$ this expectation value  becomes
\begin{equation}
\langle \xi^{2}  \rangle^e_{0,\alpha} 
=1 
+\vert \alpha \vert^{2}\tanh(\vert \alpha \vert^{2}),
\end{equation}
and therefore it turns out that when $\vert \alpha\vert \gg 1$, the expected value  $\langle\xi^{2}\rangle$ tends to $\vert \alpha \vert ^{2}$ and consequently in this limit we can say that the ``radial coordinate $\xi$ of the coherent state'' is approximated by $\vert \alpha\vert$. Note that although the particular case of $m=0$ has been taken, similar formulas apply for any $m$.

Similarly, in order to obtain information about ``the phase of the coherent state'' from the eigenvalue
$\alpha = |\alpha| e^{i\,\varphi}$, we can make use of the expression
\begin{equation}
-\xi^{2}\exp(2i\theta)=\left( \mathcal{A}^{+}-\mathcal{B}^{-}\right)^{2} .
\end{equation}
Therefore, the expectation value of this quantity in the even coherent states $\vert \Psi_{m,\alpha}^{e\pm}\rangle$ for an arbitrary and fixed $m$ is
\begin{equation}
\langle -\xi^{2}\exp(2i\theta)  \rangle_{m,\alpha}^{e\pm}
= \dfrac{\overline{\alpha}^{2}}{\cosh(\vert \alpha \vert^{2})}\left[\dfrac{1}{\sqrt{2}}+\dfrac{1}{2}\sum_{n=1}^{\infty} \dfrac{\vert \alpha \vert^{4n}}{(2n)!}+\dfrac{1}{2}\sum_{n=1}^{\infty} \dfrac{\vert \alpha \vert^{4n}}{\sqrt{(2n-2)!(2n+2)!}} \right].  
\end{equation}
From this expression, and from the  previous result (\ref{xixi}),
we can approximate $\overline{\alpha}$ by  $\xi \exp(i\theta+\pi/2)$ and it can be concluded that the relationships between the coordinate $\theta$ and and the phase $\varphi$ of the complex eigenvalue  $\alpha$
 is: $\theta+\pi /2=-\varphi$.

\subsection{Probability and current densities}

\subsubsection{ Probability density}

\begin{figure}[htb]
  \begin{center}
    \subfigure[$m=0; \,\ \vert \alpha \vert=6 \, ,\varphi=0. $]{
        \includegraphics[width=0.34\textwidth]{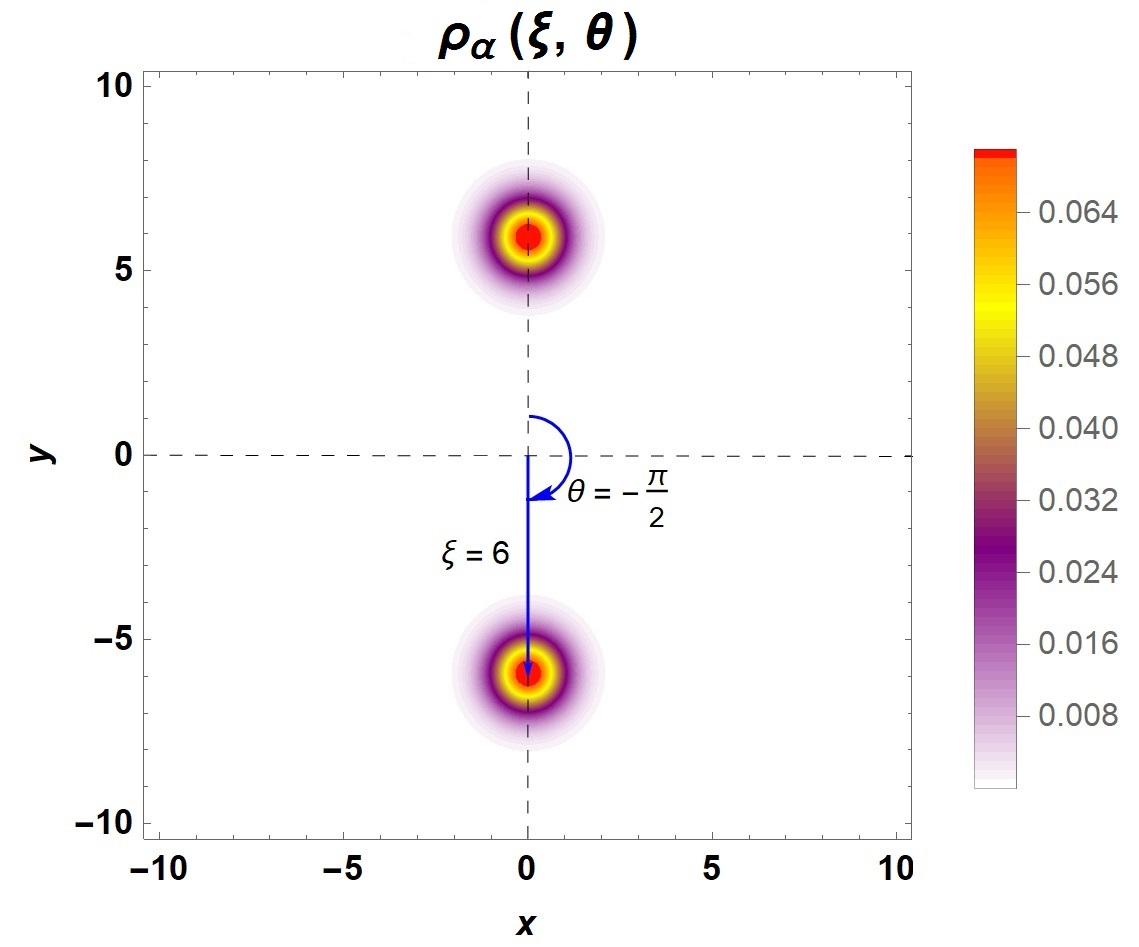}
        \label{density m=0}}\qquad
    \subfigure[$m=3; \,\ \vert \alpha \vert=6 \, ,\varphi=\pi /6. $]{
        \includegraphics[width=0.34\textwidth]{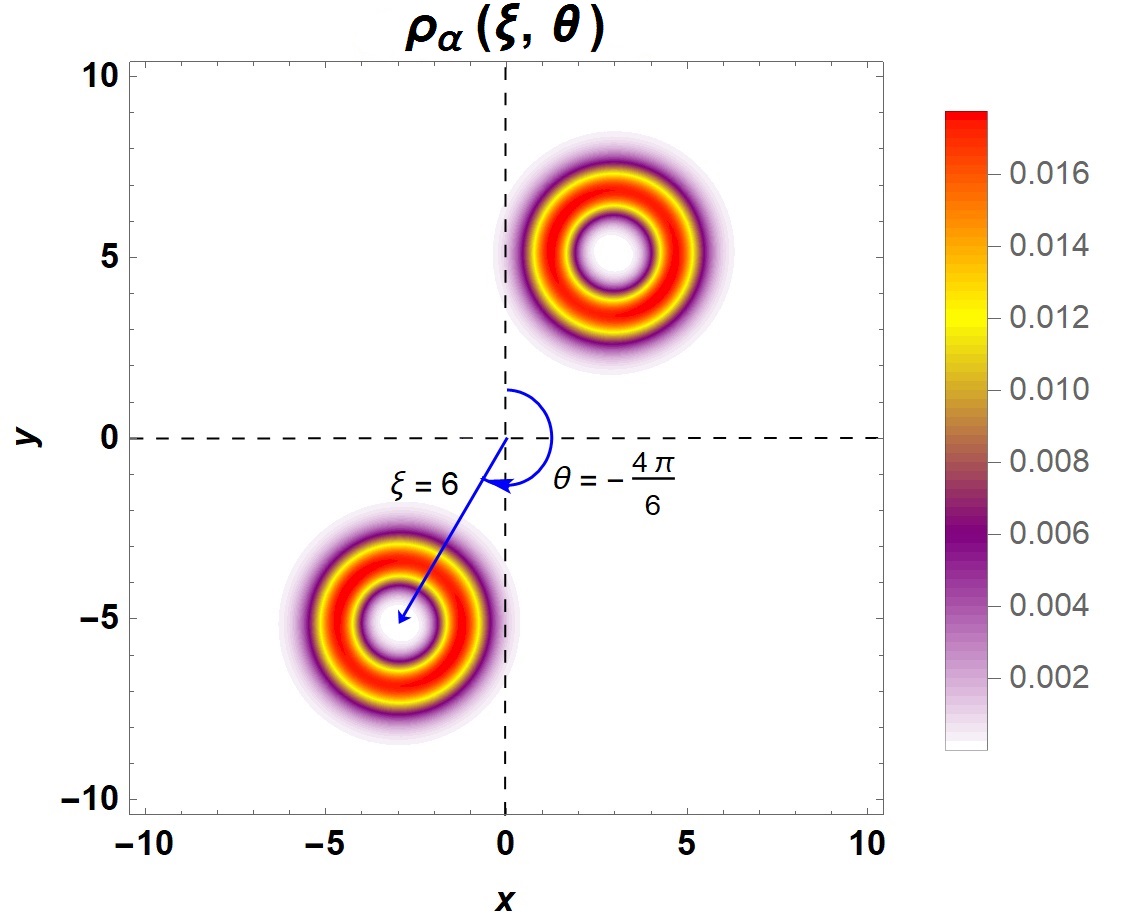}
        \label{density m=3}}\ \
  \caption{
  Probability density for the even coherent states with $m=0, 3$, taking different values of the complex number $\alpha$. To avoid too cumbersome notation, and since only even coherent states built with eigenstates of positive energy are shown, this probability density is denoted simply by $\rho_{\alpha}(\xi,\theta)$.
}
    \label{probability-densities-coherent}
  \end{center}
\end{figure}

The probability density for the coherent states \eqref{coherenteven1}--\eqref{coherentodd1} is calculated in the same way as for any other state, using \eqref{eq.rhomn}. For example, for the even coherent state this is 
\begin{equation}
\rho_{m,\alpha}^{e\pm} ( \xi,\theta)= \left( \Psi_{m,\alpha}^{e\pm} ( \xi,\theta) \right)^\dag  \, \Psi_{m,\alpha}^{e\pm} ( \xi,\theta).
\label{densityeven} 
\end{equation}
To illustrate the structure of the probability densities arising from these even coherent states, some of them are plotted in Figure~\ref{probability-densities-coherent}  for certain values of $\alpha=\vert \alpha \vert \exp(i\varphi)$ with $m=0$ and $m=3$. 
In addition, the results obtained above on the  expectation values (see subsection \ref{expecatationvalues}) are illustrated in the Figure~\ref{probability-densities-coherent}, that is, for $\vert \alpha \vert \gg1$ the value of the $\theta$ polar-coordinate of the ``position'' of the coherent state is related to the phase of the complex number $\alpha$ by $\theta=-\varphi-\pi /2$ and the $\xi$-radial coordinate of that ``position'' matches  the previous calculations, $\xi=\vert \alpha \vert$, in the same limit.

\subsubsection{Current density}

\begin{figure}[thtb]
  \begin{center}
    \subfigure[$m=0; \,\ \vert \alpha \vert=6 \, ,\varphi=0. $]{
        \includegraphics[width=0.34\textwidth]{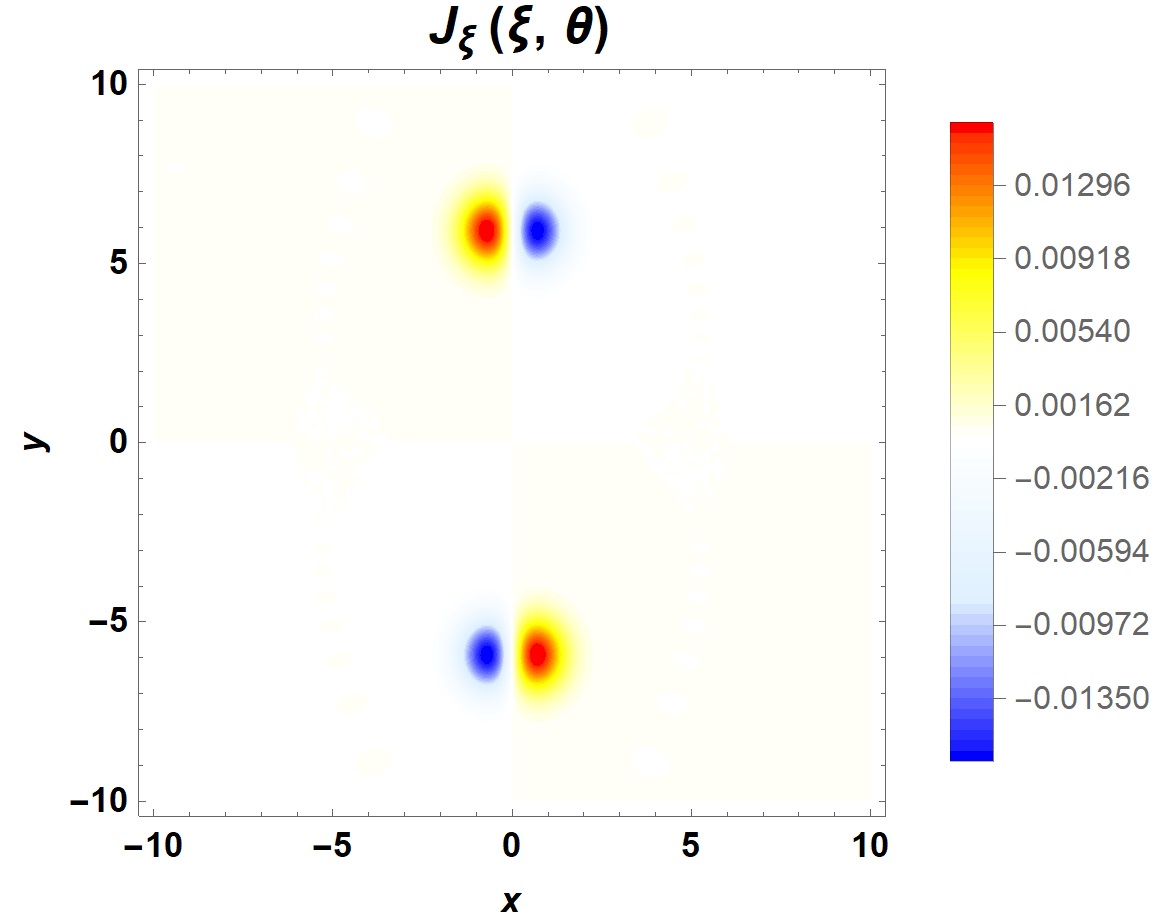}
        } \qquad 
    \subfigure[$m=0; \,\ \vert \alpha \vert=6 \, ,\varphi=0. $]{
        \includegraphics[width=0.34\textwidth]{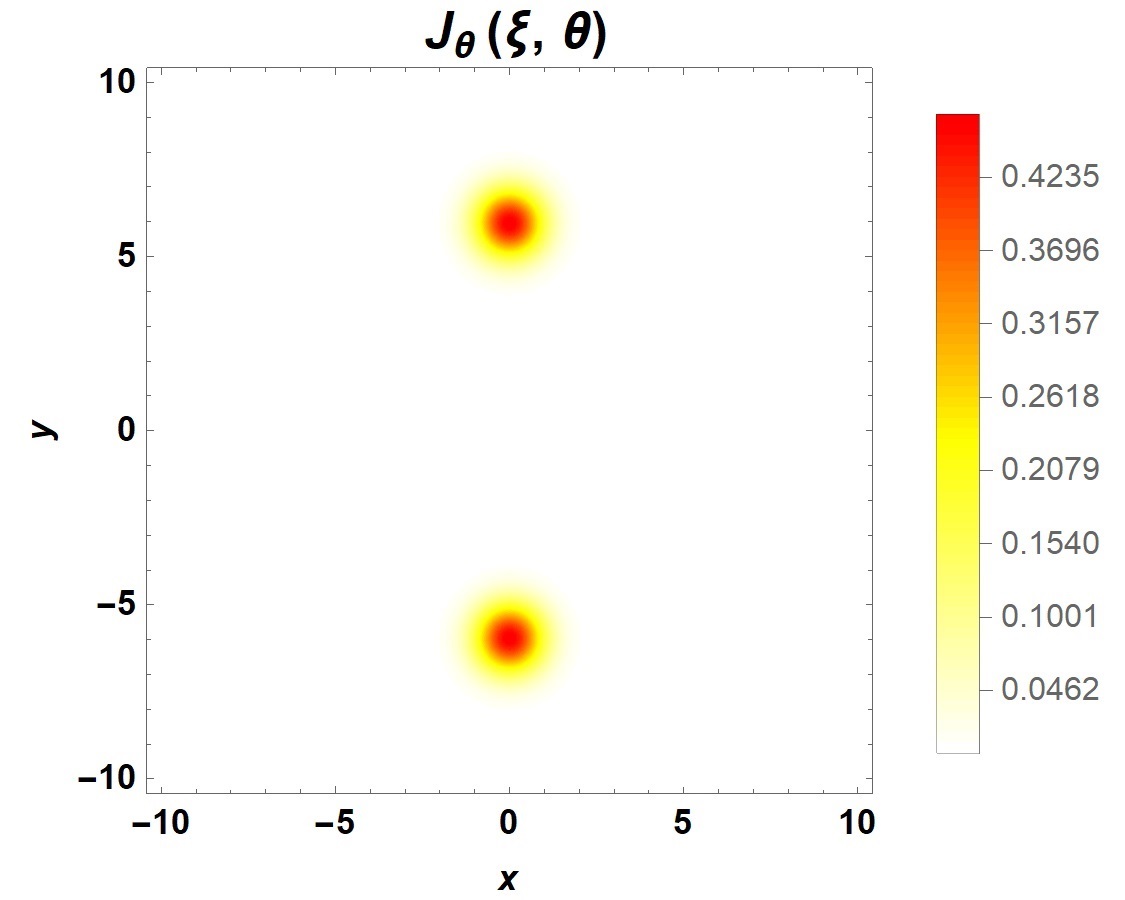}
        \label{current-density,03}}
        \subfigure[$m=3; \,\ \vert \alpha \vert=6 \, ,\varphi=\pi /6. $]{
        \includegraphics[width=0.34\textwidth]{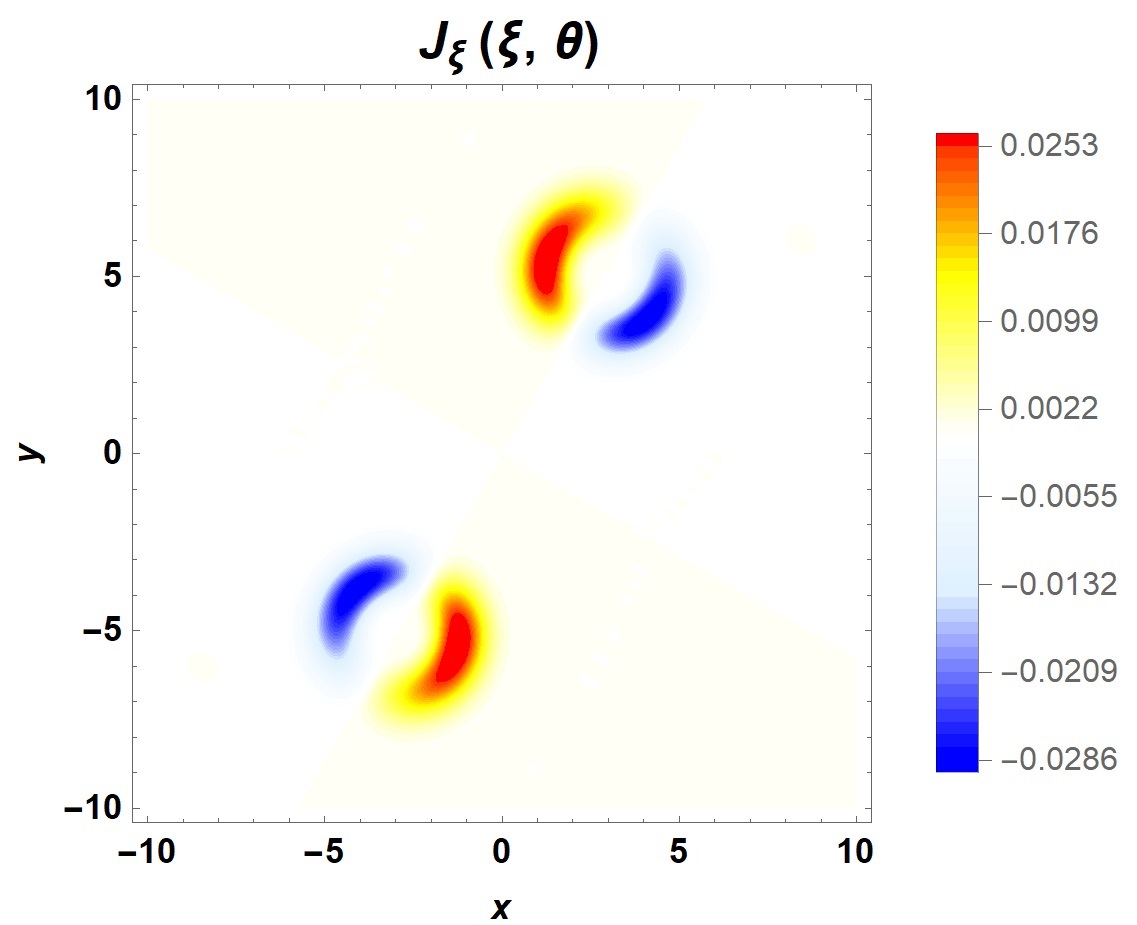}
        } \qquad
    \subfigure[$m=3; \,\ \vert \alpha \vert=6 \, ,\varphi=\pi /6.$]{
        \includegraphics[width=0.34\textwidth]{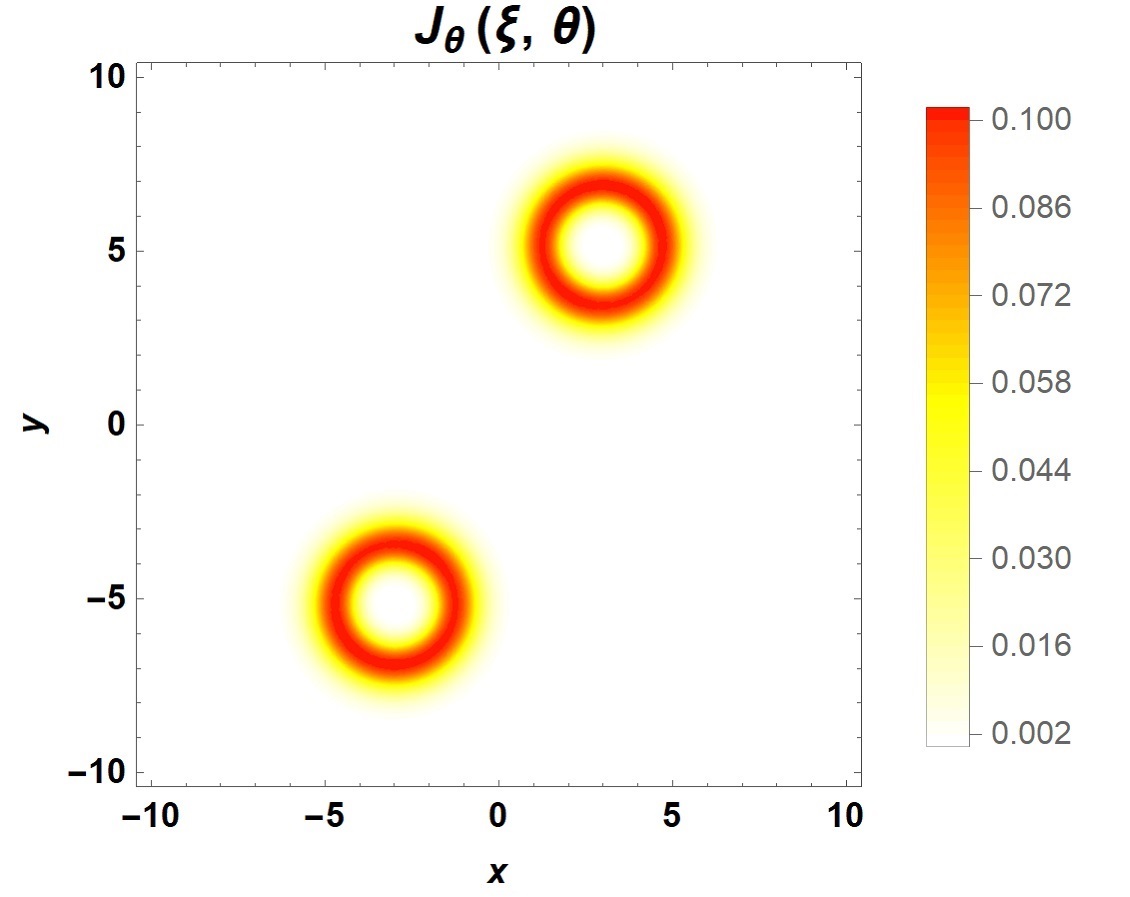}
        \label{current-density,03}}
   \caption{Radial current density $J_{\xi}(\xi, \theta)$ and  angular current density $J_{\theta}(\xi, \theta)$ for the even coherent states with $m=0, 3$, taking different values of the complex number $\alpha$.} 
    \label{currents}
  \end{center}
\end{figure}

The current densities can also be calculated for the coherent states that have been derived in this section, for which it is enough to apply the same expression \eqref{current-rtheta}. In this case we restrict ourselves to the even coherent states $\Psi_{m,\alpha}^{e\pm} ( \xi,\theta)$.
The resulting algebraic expressions are too complicated and do not give much information on their own. What is interesting is to analyze their graphic representation that better illustrates  their physical meaning. Therefore, the Figure~\ref{currents} shows the radial and angular current densities, denoted simply as $J_{\xi}$ and $J_{\theta}$, in a sufficiently illustrative specific case. The most interesting feature is the behavior of the radial component of the density current $J_{\xi}$. 
In graphs (a) and (c) of the Figure~\ref{currents}, two separate blobs are displayed with positive and negative values (red and blue colors). However, this feature is not present in the tangential component of the current density $J_{\theta}$. This rather strange representation of $J_{\xi}$ can be explained from the expression (\ref{current-rtheta}) of  both components $J_{\xi}$ and $J_{\theta}$. In these expressions 
there are  exponentials $e^{\pm i \theta}$ that causes this split into two parts. Bilayer coherent states have this property
only for the component $J_{\xi}$. However, this is not specific to bilayer graphene, it is also present in monolayer or
even in polar coherent states of non-relativistic systems in the plane.

\section{Conclusions}\label{conclusion}

This article focuses on showing how the polar coordinate symmetries of the bilayer Hamiltonian
determine many of its specific properties.
We have found three symmetries (not independent) that commute with each other: (i) the radial displacement number 
${\bf M}$, which is responsible for the displacement of the wave functions while maintaining their energy, (ii) the excitation number ${\bf Q}$ that characterizes the excitation of the states although it does not differentiates the sign
of energy, so the eigenspace with fixed eigenvalues of ${\bf M}$ and ${\bf Q}$  is two dimensional (particles and holes), and (iii) the total angular momentum ${\bf J}_z$,
which give the rotation properties in bilayer graphene and has a natural form in polar coordinates. 
As  mentioned before, this is quite important because all the magnitudes have a rotational character (whether scalar, vector or other). 
The lack of completeness of the symmetries that commute to achieve the univocal determination of a state of the system leads to a certain freedom in the definition of a coherent state, because there is no single annihilation operator. In any case, we have been able to define a reasonable type of coherent states.
Another aspect of this work that should be highlighted is
the explicit expression of the density current components in polar coordinates under a magnetic field, which was not given before.
In particular, we have shown their covariant gauge-independent character. Special interest was paid to the behavior of the radial component of the current density of a coherent state, which gives rise to a density that graphically appears separated into two drops of opposite sign.

\section{Acknowledgments}
\label{Acknowledgments}

This research was supported by Spanish MCIN with funding from European Union NextGenerationEU (PRTRC17.I1) and Consejeria de Educacion from JCyL through QCAYLE project, as well as MCIN project PID2020-113406GB-I00. DIMM was supported by CONAHCYT (Mexico), project FORDECYT PRONACES/61533/2020. DIMM acknowledges the support of CONAHCYT through the PhD scholarship 743766 and especially thanks the warm hospitality of Departamento de Física Teórica, Atómica y Óptica, Universidad de Valladolid.


\bibliographystyle{plain}





\end{document}